\documentclass{bioinfo}
\copyrightyear{2013}
\pubyear{2013}

\begin{document}
\firstpage{1}

\title[CruzDB]{CruzDB: software for annotation of genomic 
intervals with UCSC genome-browser data}

\author[Pedersen \textit{et~al}]{Brent S Pedersen\,$^{1,*}\footnote{to whom correspondence should be addressed}$,
        
         Ivana V Yang\,$^{1}$
         and
         Subhajyoti De$^{2,*}$

      }

\address{$^{1}$University of Colorado, Anshutz Medical Campus, Department of Medicine 
        12700 East 19th Avenue, 8611 Aurora, CO 80045\\
$^{2}$University of Colorado Cancer Center. 13001 E 17th Pl, Aurora, CO 80045}

\history{Received on XXXXX; revised on XXXXX; accepted on XXXXX}

\editor{Associate Editor: XXXXXXX}

\maketitle

\begin{abstract}

\section{Motivation:}
The biological significance of genomic features is often context-dependent. Annotating a particular dataset with
existing, external data can provide insight into function.

\section{Results:}
We present CruzDB, a fast and intuitive programmatic interface to the UCSC genome browser that facilitates integrative analyses of diverse local and remotely hosted datasets. We showcase the syntax of CruzDB using miRNA-binding sites as examples, and further demonstrate its utility with 3 biological discoveries. First, we find that while exons replicate early, introns tend to replicate late, suggesting a complex replication pattern in gene regions. Second, variants associated with cognitive functions map to lincRNA transcripts of relevant function. Third, lamina-associated domains are highly enriched in olfaction-related genes.

\section{Availability:}
CruzDB is available at https://github.com/brentp/cruzdb

\section{Contact:} \href{bpederse@gmail.com}{bpederse@gmail.com}, \href{subhajyoti.de@ucdenver.edu}
\end{abstract}

\section{Introduction}
Biological significance of many genomic and epigenomic features is context-dependent. Recently, large scale integrative projects such as the Encyclopedia of DNA Elements (ENCODE) project \citep{ENCODE} have systematically analyzed the regions of active transcription, gene regulation, and chromatin patterns in the genome. Even though decades of research provided insights into many individual functional elements, integrative analyses have presented a systems-level picture that could not be captured previously. Moreover, these integrative projects have highlighted that biological function of certain features can be appreciated in the context of other genomic and epigenomic features in the genomic neighborhood.
  
Systematic presentation of large-scale datasets from the ENCODE \citep{ENCODE} and other projects in the UCSC genome browser \citep{KentBrowser} has enabled individual investigators to analyze their local data in the context of these already available features. Already we are beginning to see the utility of such a community-wide integration of diverse datasets and their role in uncovering new facets of basic biology and clinical research. Researchers routinely use publicly available data-tables from the ENCODE project and many other large-scale projects from the UCSC genome browser, which also allow programmatic access to much of the information used on that site via its public MySQL servers \citep{Dreszer}. Even so, there exists no user-friendly computational framework, that allows integration of multiple in-house and publicly available data-tables and parallelized context-dependent analyses of the integrated datasets. Today, in the era of 'the \$1,000 genome, the \$100,000 analysis' \citep{Mardis}, we believe that such a computational framework can increase the speed and efficiency of integrative analyses in many areas of biomedical research. 

We present CruzDB, a programmatic interface to the genome data resources from UC Santa Cruz \citep{Dreszer} that offers a simple, parallelizable, and intuitive syntax to address common use-cases including annotation and spatial-querying. 
We first describe the design features of CruzDB, flexibility of the user-interface, and potential utilities. 
We present example code from the library and then describe four diverse findings that we made using CruzDB.

\section{Approach}

CruzDB utilizes the python programming language and sqlalchemy (SQL-alchemy) library to access publicly available data hosted at the UCSC genome browser database\citep{Dreszer} . By using sqlalchemy, we are able to wrap the database tables dynamically rather than requiring explicit code for each of the thousands of available tables (10,076 in the hg19 database). 

Although CruzDB can function using only the remote data from UCSC's MySQL instance, we show that  substantial improvements in speed can be achieved from having a local mirror, and utilizing built-in parallelization. The library contains a suite of tests to ensure correctness. CruzDB requires python 2.6 or 2.7, the MySQL client libraries and the python sqlalchemy library. Installation is available using standard python tools from http://pypi.python.org/pypi/cruzdb or from the source repository at https://github.com/brentp/cruzdb/.

\begin{methods}
\section{Methods}

CruzDB simplifies common tasks such as those that return upstream or downstream features, exons, introns, UTRs and transcription start sites. Location-based queries can utilize the UCSC bin column \citep{KentBrowser} when available for more efficient queries. The bin column that is present in some of the database tables is used to implement an efficient k-nearest neighbor search for a given feature along with methods to find nearest up and down-stream neighbors. The query results from each table can be customized, such that, for example, an interval within a CpG-island can be annotated with 'island' while one that is nearby will be annotated as 'shore'. Other operations include the generation of browser URLs to view a specific feature, the extraction of coding exons and retrieval of the genomic sequence for any of those feature types from the UCSC DAS server. One can also obtain a list of BLAT \citep{KentBLAT} hits for a particular feature.

Using CruzDB, it is possible to mirror a subset of tables from UC Santa Cruz to a local MySQL or SQLite database using a single line of python code. A local copy allows a user to add data that is not in UCSC and then use that new table just as one would any other table in the database. This expands the utility of our tool to any dataset with a start, end and chromosomal designation. Though it improves the speed of otherwise network-intensive operations, having a local copy is not necessary, and all of CruzDB's features are available on the public MySQL instance, except for those that modify the database.

In order to further speed up large numbers of queries, we provide a memory-efficient implementation of an interval tree that can be much faster than performing repeated SQL queries. Because all features must be read into memory to create an interval tree, there is a trade-off between the time to read all features into memory vs the time spent querying. That trade-off depends on the number of intervals. Figure 1 shows the comparison between local and remote instances and whether or not parallelization is used when annotating about 3,300 intervals (timing data is available in Supplementary File 1). Note that SQLite is quite fast, even without parallelization, however, the time for repeated queries to the remote (UCSC) MySQL instance can be greatly reduced by reading the entire table into a local interval tree to reduce network back-and-forth. As the number of intervals to annotate increases, so does the speed improvement from reading the intervals into a tree. Some speed improvement may be achieved by modifying MySQL settings, here we have used the default.

The most common use-case has been to annotate a list of intervals with any table from the UCSC genome-browser database. We provide an interface, by which, with a single command, a user can annotate a file of intervals with a list of tables present in the database. For gene-like tables, the output lists the nearest gene, and whether the interval overlaps an exon, intron, untranslated region, or other gene feature.
\end{methods}

\begin{figure}[!tpb]%figure1
\centerline{\includegraphics{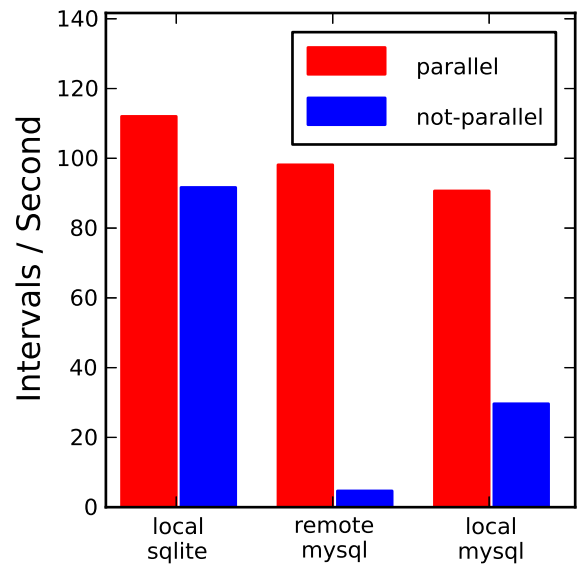}}
\caption{Intervals annoted per second for a set of about 3,300 intervals using a local SQLite,
local MySQL or remote (UCSC) MySQL instances for parallel SQL queries (red) or traditional, 
serial queries (blue).
}\label{fig:01}
\end{figure}

\section{Examples}

\subsection{Code Example: microRNA targets}

Since CruzDB is a library, we show a short code example, using the target-scan database of predicted miRNA targets \citep{Grimson} available in the UCSC genome browser as targetScanS. We will walk through the important parts of the code. The full code to perform the analysis is 12 lines (excluding comments) and is available as Supplementary File 2. First, we import the needed libraries:

\begin{verbatim}
from cruzdb import Genome
from cruzdb.sequence import sequence
\end{verbatim}

Then, we mirror the refGene and targetScanS tables from UCSC (version hg19) to a local SQLite database:

\begin{verbatim}
local = Genome('hg19').mirror(
    ('refGene', 'targetScanS'), 
     'sqlite:///hg19.mirna.db')
\end{verbatim}

Now that we have mirrored these tables from the remote UCSC server, they will always be available in the local SQLite database as long as we keep the hg19.mirna.db file. We then iterate over the rows of refGene, where each row is a python object with methods such as "is\_coding".

\begin{verbatim}
for gene in (rgene for rgene in
             local.refGene if rgene.is_coding):
\end{verbatim}

Inside that loop, we extract the gene’s 3’ UTR and search for any miRNA in targetScanS that it overlaps using the efficient bin query:

\begin{verbatim}
   utr_start, utr_end = gene.utr3
   sites = local.bin_query('targetScanS',
                           gene.chrom,
                           utr_start,
                           utr_end)
\end{verbatim}

Still inside the gene loop, we then filter to those sites that contain at least 1 miR-96 binding site with a score greater than 85 and then print those to a file along with the UTR sequence. We also save the gene name for later gene-ontology analysis:

\begin{verbatim}
    if any("miR-96" in s.name 
           and s.score > 85 for s in sites):
        print gene, sequence('hg19', gene.chrom,
                             utr_start, utr_end)
        ref_seq_ids.append(gene.name)
\end{verbatim}

After this loop, we’ll have a file of the genes that have a miR-96 binding site in their 3’ UTR. We can also send the genes to DAVID \citep{Huang} in a single command:

\begin{verbatim}
Genome.david_go(refseq_ids)
\end{verbatim}

This will open a genome browser window with the genes loaded into DAVID. Even with this short example, we identify relationships that are biologically plausible. We know that miR-96 is associated with hearing loss \citep{Mencia}; when we look at the ontology enrichment from DAVID (Supplementary File 3), we see terms associated with synapses and cell-junction which are, in turn, known to be associated with deafness and hearing loss \citep{Martinez}. While our findings in this example are not necessarily novel, it does demonstrate the utility of our approach in identifying enrichment of biologically relevant functions in the set of genes with a common miR binding site, which can be helpful in prioritizing gene lists to identify disease (or other condition) relevant regulatory elements.

\subsection{Replication Timing}
DNA replication in the human genome is spatio-temporally segregated such that some genomic regions are replicated early, and some late \citep{Hansen}. It was previously suggested that gene rich regions replicated early. But it was not surveyed whether both exons and introns replicate early, or whether the replication timing pattern is context-dependent even at a finer scale. 
Integrating DNA replication timing data from multiple cell-types, and using the definition provided by \citep{Hansen} we marked the ‘constant early’ and ‘constant late’ replication timing regions - i.e. the regions that were replicated early and late irrespective of the cell-type tested. Integrating this locally hosted dataset with CpG-island, and refGene data-tables from the UCSC genome browser, we find  that early-replicating regions are enriched for gene-bodies and for CpG-islands relative to the late-replicating regions (Supplementary Files 4 and 5), which is consistent with that reported by \citep{Hansen}. In contrast, introns were relatively more likely to be replicated late. For instance, among those regions that fall within a gene, there is 152\% enrichment for late replicating regions that fall entirely in an intron (without touching an exon) relative to early-replicating regions. When we restrict to coding genes with at least 1 intron, the enrichment goes up to 159\% (Supplementary Files 6 and 7). Although it requires further investigation, this is a novel finding that suggests that even though gene-rich regions are replicated early, there are finer-scale replication timing patterns that correlate with intron-exon structures.

\subsection{LincRNAs}
Complex genetic diseases are usually associated with multiple common and rare genetic variants. While a small subset of these variants overlap with known genes, many reside in non-protein coding regions. Some of these variants were shown to affect regulatory elements that affect expression of known genes. Non-coding RNAs (ncRNAs) are a class of regulatory RNAs that play important roles in development, cancer and other diseases. lincRNAs are a relatively recently identified class of ncRNA, which play key role in epigenetic regulation \citep{Lee}, and there are more than 20,000 predicted lincRNA genes in the human genome. So far, the genetic variants have not been systematically surveyed in the context of different classes of ncRNAs including lincRNAs. 

Here, we use lincRNA transcripts available in the UCSC hg19 from \citep{Cabili} and overlap  with the GWAS Catalog from NHGRI \citep{Hindorff} as available in UCSC's gwasCatalog table. The catalog contains a list of 12,194 SNPs that have been associated with one of over 600 traits. After annotating with CruzDB (Supplementary File 8), we examined SNPs from the GWAS catalog that overlapped a lincRNA, and especially those which were more than 10Kb from the nearest gene. Using this criteria we found 388 SNPs which overlapped a lincRNA and were also sufficiently distant from known RefSeq genes. When we enumerate the trait (disease category) with the highest proportion of SNPs that fall within a lincRNA distant to a gene and then filter to those that show at least 5 SNPs within a lincRNA, some traits among the highest by this metric are intelligence (5 out of 57 SNPs fall in lincRNAs), and other categories related to cognitive disorders (Supplementary File 9). Although overlap does not automatically indicate causality, it is consistent with the role of these miRNAs in development. There are several more instances where disease-associated variants overlap with lincRNAs with relevant biological functions. 

Using a more relaxed criteria, where a SNP was selected simply if it was closer to a lincRNA than to the nearest gene, we found 2153 SNPs (Supplementary File 10). Our findings, combined with the recent study showing a lower incidence of SNPs within lincRNAs \citep{Chen} show the importance of annotating GWAS results with lincRNAs in addition to genes.

\subsection{Lamina Associated Domains}
Within the nucleus, different genomic regions occupy distinct nuclear territories, such that some regions are in contact with nuclear lamina (termed lamina-associated domains or LADs) \citep{Guelen,Dittmer}. These regions usually have repressive chromatin marks and lower levels of gene expression. However, it has not yet been investigated systematically whether certain classes of genes are more clustered in LADs compared to that expected by chance. Overlaying data on lamina associated domains (LADs) from \citep{Guelen}, and known genes, we find over 5000 genes overlap completely/partially with the LADs (Supplementary File 11). Furthermore, piping the genes that overlap a LAD with a score \textgreater 0.9 (the fraction of probes with a positive smoothed log-ratio) to the DAVID gene-ontology enrichment software \citep{Huang} we report very strong enrichment for categories related to olfaction (adjusted p \textless 1e-80), G-protein coupled receptor (adjusted p \textless 1e-60), and other categories related to sensing (Supplementary File 12).  Our findings are consistent with a recent report \citep{Clowney} that nuclear clustering of olfactory receptor genes governs their monogenic expression. It is suspected that laminB receptor-induced changes in nuclear architecture influences singular transcription pattern of the olfactory receptor genes \citep{Clowney}.

Furthermore, when we filter to genes that are strictly contained within a LAD (not merely overlapping) with a score \textgreater 0.9, and send that stricter subset of 2,570 genes to DAVID, we find even stronger enrichment of olfaction and related terms (adjusted p \textless  1e-106), g-protein coupled recepter (adjusted p \textless  1e-95) (Supplementary File 13).

%\begin{figure}[!tpb]%figure2
%\centerline{\includegraphics{fig02.eps}}
%\caption{Caption, caption.}\label{fig:02}
%\end{figure}

\section{Discussion}

We have introduced CruzDB, a parallelizable and intuitive syntax-based programmatic interface with UCSC genome browser that allows integrative context-dependent analyses of diverse local and remotely hosted datasets, as well as annotation and spatial-querying. Some of the functions that make CruzDB a library of broad and general utility are the feature extraction, fast queries, and simple syntax.
Using the library, one can mirror the UCSC databases to a local SQLite or MySQL database, perform location-based queries, and perform integrative analyses combining local and remotely hosted features. We have shown how to create a local copy of selected tables is a single line of code and how having that local copy improves the speed of later analyses.

We showcase the programmatic interface of CruzDB using miRNA-binding sites as examples, and further demonstrate its utility using 3 biological examples, each with a potentially novel discovery.
First, we showed the syntax of the library by extracting genes with a target site for the miR-96 microRNA. 
Second, by integrating exon and DNA replication timing data, we show that even though exons typically replicate early, introns are likely to replicate late during S phase. 
Our findings suggest a more complex DNA replication landscape than previously appreciated. 
Third, although current GWAS studies have primarily focused on functional variants affecting protein-coding genes, some variants are likely to affect other functional elements including non-coding RNAs. 
We report several instances where disease-associated variants overlap with lincRNAs with relevant biological functions. 
For example those related to intelligence and cognitive disorders. 
Our findings, combined with the recent study showing a lower incidence of SNPs within lincRNAs \citep{Chen}, highlight the importance of examining GWAS hits in this context. 
Finally, integrating data on lamina-associated domains and protein-coding regions, we find that olfactory receptor genes are highly enriched in the lamina-associated domains. Our findings are consistent with a recent report that nuclear clustering of olfactory receptor genes governs their monogenic expression \citep{Clowney}. It is suspected that laminB receptor-induced changes in nuclear architecture influence singular transcription pattern of the olfactory receptor genes \citep{Clowney}. Further work needs to be done to demonstrate the broader impact of our findings in each of these four biological cases in detail, we aim to pursue them outside the scope of this method paper. Nevertheless, the four examples outline the broad utility of CruzDB, and its applications in diverse areas of biomedical research. 

%%%%%%%%%%%%%%%%%%%%%%%%%%%%%%%%%%%%%%%%%%%%%%%%%%%%%%%%%%%%%%%%%%%%%%%%%%%%%%%%%%%%%
%
%     please remove the " % " symbol from \centerline{\includegraphics{fig01.eps}}
%     as it may ignore the figures.
%
%%%%%%%%%%%%%%%%%%%%%%%%%%%%%%%%%%%%%%%%%%%%%%%%%%%%%%%%%%%%%%%%%%%%%%%%%%%%%%%%%%%%%%

%\section*{Acknowledgement}

%\paragraph{Funding\textcolon} 

%\bibliographystyle{natbib}
%\bibliographystyle{achemnat}
%\bibliographystyle{plainnat}
%\bibliographystyle{abbrv}
%
%\bibliographystyle{plain}
%\bibliographystyle{bioinformatics}
% \bibliographystyle{natbib}

%\begin{thebibliography}{}
\bibliographystyle{natbib}
  \bibliography{cruzdb}
%\end{thebibliography}
\end{document}